\documentclass[runningheads]{llncs}
\usepackage{graphicx}
\usepackage{amsmath}
\usepackage{amsfonts}
\usepackage{tikz}
\usetikzlibrary{shapes,arrows}

\begin{document}
\title{A Perspective on the Ubiquity\\of Interaction Streams in Human Realm}
%
%
\author{Damian Serwata\orcidID{0000-0003-1509-4495} \and
Mateusz Nurek\orcidID{0000-0003-1863-9331} \and
Rados{\l}aw Michalski\orcidID{0000-0002-0106-655X}}
\authorrunning{D. Serwata et al.}
%
\institute{Wroc{\l}aw University of Science and Technology\\Department of Artificial Intelligence\\Wybrze\.ze Wyspia\'nskiego 27\\50-370 Wroc{\l}aw, Poland\\
\email{\{damian.serwata,mateusz.nurek,radoslaw.michalski\}@pwr.edu.pl}}
%
\maketitle              
\begin{abstract}
Typically, for analysing and modelling social phenomena, networks are a convenient framework that allows for the representation of the interconnectivity of individuals. These networks are often considered transmission structures for processes that happen in society, e.g. diffusion of information, epidemics, and spread of influence. However, constructing a network can be challenging, as one needs to choose its type and parameters accurately. As a result, the outcomes of analysing dynamic processes often heavily depend on whether this step was done correctly. In this work, we advocate that it might be more beneficial to step down from the tedious process of building a network and base it on the level of the interactions instead. By taking this perspective, we can be closer to reality, and from the cognitive perspective, human beings are directly exposed to events, not networks. However, we can also draw a parallel to stream data mining, which brings a valuable apparatus for stream processing. Apart from taking the interaction stream perspective as a typical way in which we should study social phenomena, this work advocates that it is possible to map the concepts embodied in human nature and cognitive processes to the ones that occur in interaction streams. Exploiting this mapping can help reduce the diversity of problems that one can find in data stream processing for machine learning problems. Finally, we demonstrate one of the use cases in which the interaction stream perspective can be applied, namely, the social learning process.

\end{abstract}
\section{Introduction}
To describe reality, we often use constructs that help us to make it abstract. Pythagoras, who is attributed with the belief that \textit{all things are number}, claimed that numbers are the perfect representation of reality and -- actually -- reality is numbers. Following up on this philosophy, Shakuntala Devi, a writer and mental computer, twenty-five centuries later said that ,,\textit{Without mathematics, there's nothing you can do. Everything around you is mathematics. Everything around you is numbers}''. In fact, numbers are one of the most universal concepts to use when describing reality. However, numbers are not enough. Even leaving aside the artistic point of view, where poets, painters and writers may have a completely different take on the means of describing reality, we could look for other representations rooted in mathematics. One of these is 
\textit{networks}, i.e. sets of linked entities, where each entity is labelled and means someone or something, and that link also has a meaning. In fact, shifting the Pythagoras' perspective, network scientists claim that \textit{networks are everywhere}~\cite{posfai2016network}. To give only one example, related to human beings, from cells through the human brain, body, individuals and groups, ending up with whole societies -- everywhere there we can find links between objects, and exploiting these can lead to a better understanding of the complexity of the world in micro--, meso--, and macroscale.

However, in both cases, \textit{numbers} and \textit{networks}, one needs to remember that these constructs (and many more) are a \textit{representation}, not the \textit{object} itself. And as such, we rely on an approach that might be good in describing given object in some cases, but completely fails in others.

In this work we are mostly interested in a social context, so we focus on answering the question how to accurately model interactions of people~\cite{hare1965small}, diffusion of information~\cite{lerman2010information} and innovations~\cite{rogers2014diffusion}, spread of influence~\cite{kempe2003maximizing}, or social learning~\cite{barkoczi2016social}. Albeit the work itself does not provide an experimental evidence for our arguments, we argue that in the aforementioned contexts we should be more open to a certain approach that is linked to how we acquire and parse information, i.e. interaction streams. Yet, before providing the justification for this approach, we would like to take a step back and investigate how the modelling of these processes looked like from the chronological perspective, which was also related to the increasing complexity of this task.

This work is structured as follows. In the next section we provide a historical background on modelling human interactions. Section~\ref{sec:data-streams} describes the mechanics of data streams, which -- in our perspective -- should be thought of a relative of interaction streams. Section~\ref{sec:analogies} explains this relationship and provides more justification for exploring this direction as the best candidate for further studies.  In Section~\ref{sec:perspectives} we take a perspective on stream modelling. In order to provide one use case in which we demonstrate the capabilities of the proposed approach, we look at the social learning phenomenon. This use case is investigated in Section~\ref{sec:social-learning}. Lastly, Section~\ref{sec:summary} concludes this work.

\section{Modelling Social Interactions}
\label{sec:modelling-social-interactions}
Firstly, we have to ask ourselves the question on how we can formally think of a social interaction. In its basic nature, this can be thought of a tuple $iv^e_{ijk} \in IS $ (interaction sequence) presented in Eq.~\ref{eq:social-interaction-tuple}, where $v^e_i$ and $v^e_j$ represent individuals from a set of all individuals $V$ ($v^e_i$ and $v^e_j \in V$), with the requirement that $v^e_i \ne v^e_j$ and $t_k$ is a discrete timestamp of the interaction, e.g. time in which the interaction took place.

\begin{equation}
\label{eq:social-interaction-tuple}
iv^e_{ijk} = (v^e_i, v^e_j, t_k)
\end{equation}

This can be considered the simplest form of representing an interaction, yet instantly one can think of its variants. For instance, if there are multiple types of interactions, the tuple presented in Eq.~\ref{eq:social-interaction-tuple} can be extended by the interaction type. Next, if we consider the duration of interactions, this can become an additional element of the tuple.

One important remark needs to be made in the context of how we interact. Assuming pairwise synchronous interactions we can say that $v_i$ can be swapped with $v_j$. If this is not the case, the ordering needs to be kept and possibly the sending and receiving times need to be distinguished. Lastly, the requirement that people interact pairwise can also be relaxed by extending the tuple to the form presented in Eq.~\ref{eq:social-interaction-simplex}. 

\begin{equation}
\label{eq:social-interaction-simplex}
iv^e_{ijk} = (V^e_{in}, t_k)
\end{equation}

Here, $V^e_{in}$ is a set of interacting individuals, and $V^e_{in} \in V$. This concept will be further explored when we will be discussing the simplicial complexes or hypergraphs.

Social networks are one of the most widely used concepts for deriving knowledge on top of these interactions~\cite{wasserman1994social}. They are a form of aggregating social interactions into a more complex structure that allows to investigate how individuals are interconnected. Yet before thinking of collapsing social interactions in one structure, it might be beneficial to study ego networks~\cite{dejordy2008introduction}, i.e. the networks that have a single individual as the root and its alters as direct neighbours. To convert social interactions into an ego network one needs to filter the set of tuples all presented in Eq.~\ref{eq:social-interaction-tuple} to a specific $v_i$. As a result we are going to have this \textit{ego} as a centre of the network and all its \textit{alters} as neighbours.

Widening the scope of the analysis to all interacting individuals unveils the concept of social networks. A social network is a tuple $ SN = (V, E) $, where $ V = \{v_1, \ldots, v_n \}, n \in \mathbb{N}_+ $ is the set of vertices and $ E = \{ e_1, \ldots, e_{k^e} \}, k^e \in \mathbb{N}_+ $ is the set of edges between them. Each vertex $ v_i \in V $ represents an individual $ v_i^e $ from social interactions and each edge $ e_{ij} $ corresponds to the directed social relationship from $ v_i $ to $ v_j $, such that $ E = \{ (v_i, v_j, w_{ij}) : v_i \in V, v_j \in V, v_i = v^{e}_i, v_j = v^{e}_j $ and $ \displaystyle\mathop{\forall}_{ij}(\displaystyle\mathop{\exists}_{k} iv_{ijk} \in IS \Leftrightarrow e_{ij} \in E), w_{ij} \in [0,1] \} $. Here, value $ w_{ij} = \frac{n^e_{ij}}{n^e_{i}} $ denotes the importance (weight, strength) of the relationship between individuals, such that $ n_{ij}^e $ is the number of events $ iv_{ijk} $ from $ v^e_i $ to $ v^e_j $ in $ IS $ (regardless $ k$) and $ n_{i}^e $ is the number of all events initiated by $ v_i^e $ (outgoing from). Note that despite the fact that both $v$ (vertices) and $v^e$ (individuals, see Eq.~\ref{eq:social-interaction-tuple}) belong to $V$, their interpretation is slightly different, so we decided to distinguish them by using an upper index~$e$.

In the definition above, we assumed a directed social network, but relaxing this assumption is also quite often made, it all depends entirely on the context.

Social networks became a very useful tool for mapping social interactions to a bigger landscape. Yet, whenever one builds such a social network on top of social interactions, a number of decisions have to be made. For instance, whether all interactions shall be collapsed into a single social network, resulting in a static network, or if we are interested in keeping a semi-ordering of these by using temporal networks~\cite{holme2012temporal,michalski2020entropy}. Similarly, the decision needs to be taken if all types of events are made equal or we distinguish them, resulting in multigraphs~\cite{royle2006graphs} or multilayer networks~\cite{kivela2014multilayer,stepien2023role}. These network models can be further supported by using data assimilation methods and machine learning to build a structure that better corresponds to reality~\cite{lever2023human,cheng2024real}, but this adds another level of complexity.

Going back to the definition of a social interaction linking more than two individuals (see Eq.~\ref{eq:social-interaction-simplex}), this can also be represented in social networks by either simplicial complexes~\cite{salnikov2018simplicial} or hypergraphs~\cite{berge1984hypergraphs}. In both cases, we link more than two vertices and as such we are able to represent interactions that are multi-party. Yet, in this case, one needs to take into account that the whole apparatus developed for static networks needs to be rethought, as common measures and metrics will not be relevant anymore.

Given that many social networks tend to be large, it is sometimes impossible to find optimal solutions for certain challenges and heuristics are often used, e.g. influence maximisation~\cite{michalski2015maximizing,weskida2019finding}. Another line of simplifying the problem is to use network embeddings, so that graphs can be potentially expressed as fixed-length vectors in order to be processed in a machine learning pipeline~\cite{grover2016node2vec,torricelli2020weg2vec}. Obviously, each approach has its own advantages and disadvantages, yet these methods have become very popular in recent years as they have opened the possibility of achieving at least approximate results for certain problems.

Yet, when chasing for these solutions, we tend to forget how many layers of abstraction are built on top of social interactions. To give the reader an impression of that, in Figure~\ref{fig:pipeline} we present one of the modern pipelines that is used for the link prediction task~\cite{cao2019network}.

\tikzstyle{block} = [draw, fill=blue!20, rectangle, 
    minimum height=3em, minimum width=1em]
\tikzstyle{input} = [coordinate]
\tikzstyle{output} = [coordinate]
\tikzstyle{pinstyle} = [pin edge={to-,thin,black}]

\begin{figure}
\centering
\begin{tikzpicture}[auto, node distance=2.5cm,>=latex']
    \node [block, name=soc-int] (soc-int){\shortstack{Social\\ interactions}};
    \node [block, right of=soc-int] (filtering) {\shortstack{Filtering\\ data}};
    \node [block, right of=filtering] (building) {\shortstack{Building\\ network}};
    \node [block, right of=building] (embedding) {\shortstack{Network\\ embedding}};
    \node [block, right of=embedding] (prediction) {\shortstack{Link\\ prediction}};
 \draw [draw,->] (soc-int) -- node {} (filtering);
 \draw [draw,->] (filtering) -- node {} (building);
 \draw [draw,->] (building) -- node {} (embedding);
 \draw [draw,->] (embedding) -- node {} (prediction);
\end{tikzpicture}
\caption{An exemplary pipeline used for the link prediction task.}
\label{fig:pipeline}
\end{figure}
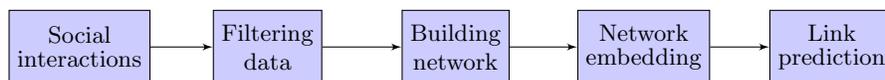
What we see here is that the path leading from social interactions to the actual task that relates to the original concept (as we want to predict if people will interact) covers a number of steps in which multiple design decisions need to be made. And given that an error made at an earlier stage propagates further, so the carefulness is crucial here.

To show that this can happen quite often, we chose a number of examples related to the third step of Figure~\ref{fig:pipeline}. For instance, in~\cite{holme2012temporal} authors underline that using time windows can introduce significant differences, since crucial moments (e.g. time-sensitive sequences of events) can be flattened into one snapshot leading to unpredicted results of the spreading outcomes. This has also been studied in detail in~\cite{karsai2011small}. In~\cite{saganowski2012influence} authors investigated whether the group evolution prediction task is dependent on the window type and size eventually confirming this hypothesis. Similar conclusions have been drawn in the area of social influence maximisation~\cite{michalski2020effective,michalski2015maximizing}. This sample of works demonstrates that in a long chain of design decisions the final outcomes can vary depending on the modelling applied.

Further in this work we would like to base on the argument of the complexity of a current modelling process to urge for stepping down from this path and to stay closer to interactions understood as interaction streams. Moreover, as these interaction streams relate to data streams, our goal is to also show that it might be possible to leverage the techniques known in data stream mining to reduce negative effects that often relate to social phenomena. However, before building the bridge between interaction streams and data streams, we need to provide some basic facts about them, and this takes place in the next section.

\section{Stream Data Mining}
\label{sec:data-streams}
The modern, digital era has abounded in an unprecedented volume, velocity, and variety of data leading to ongoing transformation of present society. There are multiple applications in computational sciences domain in which massive datasets processing is essential. Many of those data sources are related to human activity. Records describing our actions are produced actively or passively by software and hardware that we interact with on a daily basis. The process of data generation and consumption changed and enforced a new paradigm of data stream mining to be introduced for exploration and learning from such kind of data. In response to the dynamic and continuous nature of data streams, traditional static dataset processing methods were proved to be insufficient.

Data streams can be formally defined as an ordered sequence of data records: $S = \{s_1, ..., s_n\}$, where the size of the sequence $n(S)$ can be infinite~\cite{rutkowski2019stream}. Each element of the data stream denoted as $s_x$ can be unpacked as a tuple of attributes belonging to this element. The set of attributes depends on the specific context of the data stream and might involve features like timestamp, duration, type, content, etc.

There are three essential properties of data streams resulting in the need for adjustments in the well-established data processing methods:
\begin{itemize}
  \item Massive and possibly infinite number of incoming data records.
  \item Rapid arrival of data reaching the system.
  \item Modifications in the distribution of data in time.
\end{itemize}
The volume of data makes it impossible to approach it from the static dataset perspective as its infinitely growing size prevents it from being collected, stored and ultimately processed. High velocity demand algorithms capable of processing data at the moment of data record arrival. Non-stationary distribution of the data, which is called concept drift, underscores the importance of adaptive nature of models providing possibility of dynamic adjustment to changing patterns.

The unique characteristics of data streams makes them impossible to be processed using most of the classical batch-processing machine learning algorithms. Those features need to be included into the notion of appropriate stream data mining algorithms. A number of methods were developed over the years to address those requirements. Examples of algorithms applicable in the streaming data context are Bayesian models, artificial neural networks, and properly adjusted tree-based algorithms. Working with data streams often requires the application of preprocessing techniques before learning can be performed. These procedures involve data-based techniques including sampling, aggregations, synopsis data structures, sketching, load shedding, and embeddings as well as task-based techniques such as sliding windows, algorithm output granularity, and approximations.

Stream data mining finds applications across diverse domains, including finance~\cite{lin2018using}, healthcare~\cite{zhang2012real}, telecommunications~\cite{weiss2010data}, and environmental monitoring~\cite{cabrera2018environmental}. In social sciences a dominant approach, that was proven to be useful for modelling social environments in different kinds of collective adaptation processes was based on the social network concept. Here, we advocate that looking at social science models through the lens of data streams would be beneficial for both closer modelling and a better understanding of various social processes and development of more efficient methods of data stream preprocessing based on cognitive and social mechanisms.

\section{At the Intersection of Social Interactions and Data Streams}
\label{sec:analogies}
Research on rules driving social life has been conducted in many disciplines and from various perspectives. Despite different types of complex processes e.g. social learning, social influence, or belief dynamics taking place in societies, they all can be conceptually expressed and later studied within one unified framework of collective adaptation~\cite{galesic2023beyond}. All three co-evolving building blocks of the framework (social integration strategies, social environments, and problem structures) depend and develop based on social interactions. Those social interactions are essentially the driver of all social processes arising in our world. The adaptation occurring in collectives, regardless the social process taking place, requires communication between entities constituting these collectives. It is because of interactions and as their result, an individual's internal state changes – whether it involves the evolution of one's attitudes, opinions, beliefs, or influence. Having defined the data streams and the perspective on modelling social structures of collectives, we can now investigate the applicability of such a joint approach to represent elements of complex social systems. We want to focus on interactions occurring between individuals themselves and between individuals and other entities, and the way how these interactions are perceived.

The streaming approach is a natural framework that people leverage in everyday life for processing signals incoming from their environments. From a cognitive perspective, events are a fundamental way of perceiving and experiencing for humans. Events emerge in our minds through the discretization of continuous stimuli flowing from the surrounding reality. Events are an inherent element of our cognition, helping us understand the world and predict the future~\cite{radvansky2011event,radvansky2014event}. Social interactions are no exception, they are also received and processed internally as a sequence of events. This intuition motivated us to examine various parallels between the conditions and details of data processing by people and streaming algorithms. One example is arriving of the high volume of data and its surrounding circumstances. Another aspect is the high speed of incoming signals and the following requirement to respond to them quickly. Both the variety and quality of received records form another viewpoint. Finally, the dynamic nature of the surrounding environments is also present in both contexts.

Further in this section, we present analogies between social interaction and data streams (Table~\ref{tab:parallels}); however, the list of provided parallels is certainly not an exhaustive one. More in-depth analysis is likely to reveal further similarities. Nevertheless, the presented comparisons offer intuition and justification to consider social interactions as data streams in modelling collective behaviour.

\subsection{Volume -- What to Remember When Overloaded}
People, similarly to stream mining algorithms, have limited memory, that is not capacious enough to store all the signals incoming for the environment. The information overload describes a situation when people struggle to process and absorb a large volume of information~\cite{Bawden_Robinson_2020,holyst2024protect}. When overwhelmed, people tend to rely on unconscious and automatically applied heuristics, which result in the presence of cognitive biases~\cite{Azzopardi_2021}. Availability bias appears as people give greater attention to the information that was already observed~\cite{Tversky_Kahneman_1974}. Anchoring bias happens when people overestimate the importance of the first information, even though it might be incorrect. Lastly, when facing too much information, people tend to focus on those records that are aligned with their assumptions made upon prior observations~\cite{Nickerson_1998}. The design of streaming algorithm also accounts for dealing with a high volume of data. Sampling is a preprocessing technique for data stream mining used to select only a small part of the stream for computation. Sampling methods aim to provide a maximally representative subset of data records. The described cognitive heuristics for the selection of incoming signals are analogous to different types of sampling strategies. 

Another aspect is, that the information about past records needs to be stored in a more compact way. The items that are assessed as important and worth to remember cannot be simply stored to be retrieved later as needed, because of the memory restrictions. To be useful, they have to be included into an individual's knowledge. This kind of reduction may results in order effects like primacy or recency effects. These effects appear when the specific position of collected information results in a greater or lesser impact on later actions~\cite{Bansback_Li_Lynd_Bryan_2014}. A reference to the mentioned effects might be found in use of aggregations for processing data streams. The role of aggregation is to represent the number of elements in a more compact form, that still preserves information about the elements collection. Descriptive statistics, like average or maximum value are examples of aggregations. Various aggregations might be more or less accurate for different cognitive processes.

\subsection{Speed -- Need for Action Under Time Pressure}
Often, people must act upon received information immediately. For decision making, too much information present at the same time leads to abiding by better known options, even if they are less effective~\cite{Frisch_Baron_1988}. Decoy effect, happens when a new information about choice possibility causes change in decision between previously considered options~\cite{Huber_Payne_Puto_1982}. New signals, that should be irrelevant for the cognitive process, affect it in a situation when there is too little time to carefully analyse all related data. Likewise, for stream mining, the pace of incoming records varies in time and processing time is expected to be as short as possible, and especially no longer than some maximal bound. One adaptive approach addressing this issue is algorithm output granularity method~\cite{gaber2003adaptive}, that manages high and unstable data rates and adjusts to available processing time and memory.

\subsection{Variety and Quality -- When Knowledge Is Incomplete}
In reality people rarely act upon full information. Instead, we have to mitigate the effects of incomplete knowledge. Sometimes the part information that we receive requires us to estimate the whole distribution of it, e.g. we need to anticipate what was the whole content to make any actions or at least educated guesses on how to act. In this case one can think of variety of parallels to data streams, since we can recreate the distribution of information based on previous occurrences of similar data. Or, on the other hand we can apply other data filling strategies, such as the last observation carried  forward~~\cite{lachin2016fallacies}. Moreover, the information we receive can be noisy or biased, either because of natural disturbances of the system or because it went through other people who converted it. In these situations we are also capable of filter noise or recreate original information~\cite{peelle2022our}.

\subsection{Concept Drift -- World's Dynamic Nature}
Environments that we are embedded in are naturally dynamic. The complex interactions between people and other entities effect in changes in the characteristic of individual agents and collectives. The relations between individuals, usually represented as links in a network, play a pivotal role. As these relationships evolve, opinions are exchanged, and information spreads within the network, influencing both the nodes and the topology of the network itself. Changes in social ties can lead to shifts in adopted beliefs, the frequency of interactions, and the strategies embraced by individuals. People are able to detect and adapt to some of these changes easily, and others might pose multiple problems. For example, leveraging some exploitative strategy might result in a change of the social environment of an individual and fewer social interactions in the future. Another case might be an invalid vaccination policy against a virus, or overuse of antibiotics against some bacteria, that may eventually cause their rapid evolution and development of resistance to utilized assets. Streaming algorithms design, likewise, aim for capability for such data characteristic change and adaptation to the new reality.

\begin{table}
\centering
\scriptsize
\caption{Examples of parallels between stream data mining concepts and social phenomena}
\label{tab:parallels}
\begin{tabular}{|p{0.15\textwidth}|p{0.15\textwidth}|p{0.7\textwidth}|}
\hline
\textbf{Stream Data Mining Concept} & \textbf{Social Phenomenon} & \textbf{Linkage between these two concepts}\\
\hline
Data Volume         & Information Overload          & High volume of incoming data can overwhelm individuals, leading to difficulty in processing and making sense of information.\\ \hline
Data Speed          & Action Under Time Pressure    & Rapid arrival of data requires quick decision-making analogous to the need for timely actions in fast-paced social environments. \\ \hline
Data Variety        & Incomplete Knowledge          & Comprehensive understanding is challenging with diverse sources and varying data quality much like the individuals' incomplete knowledge due to diverse perspectives, credibility and limited information. \\ \hline
Data Quality        & Misinformation and Fake News  & Ensuring the accuracy and reliability of data parallels the challenge of facing and mitigating misinformation and fake news in social contexts, where ensuring the credibility of information is crucial. \\ \hline
Concept Drift       & World’s Dynamic Nature        & Continuous changes in data patterns necessitate adaptive models and strategies, similar to how social environments evolve over time, adjusting to the present conditions. \\ \hline
Scalability         & Society Growth                & The ability of data mining algorithms to handle increasing amounts of data reflects the opportunities posed by growing populations, such as increased collective intelligence capabilities. \\ \hline
Model Interpretability & Understanding Social Behavior & The interpretability of data mining models reflects the need to understand complex social behaviors and dynamics. \\ \hline
Data Imbalance      & Social Inequality             & Addressing data imbalances in stream data mining mirrors efforts to tackle social inequalities, where ensuring fairness and equity in data representation is crucial for promoting inclusivity and social justice. \\ \hline
Anomaly Detection   & Abnormal Behavior             & Detecting anomalies in data streams reflects the identification of abnormal behavior in social contexts, where recognizing deviations from norms is important for maintaining order and security. \\
\hline
\end{tabular}
\end{table}


\section{Streaming Modelling Perspectives}
\label{sec:perspectives}
We believe that bringing streaming perspective into the realm of modelling collective adaptation phenomena on top of social interactions has the potential to both deepen our understanding of consuming social interactions, and improve social processes models' design by bringing them closer to the real-world scenarios. It can be especially beneficial for settings limited by inability to operate on aggregated models of social environments like social networks. 
This approach of looking at social processes through the prism of data streams could be viewed from an even broader angle for investigation of various collectives that perform some kind of interactions, at different levels of complexity, starting from inanimate matter, through mono-cellular, and multi-cellular organisms and organs, up to the level of animals, societies and ecosystems. In this work we are dedicated to analysis of human society only. In particular, we propose two distinct perspectives for leveraging the streaming approach, at the level of a single individual and the level of a multi-agent collective. Some of the properties of the social interaction stream are maintained for both of them. A single individual has to be involved in the interaction as a source or a target. 

Social interaction sequence can be directly considered as an instance of a data stream, so $S = IS$. Then a single interaction is equal to a single data record registered in the stream: $s_x = iv^e_{ijk}$. A timestamp as well as individuals involved in the interaction are attributes of the data record in the streaming context. Here, we do not consider other attributes of interaction, although including additional information describing an event would be most likely beneficial for modelling certain phenomena. It is worth noting that different types of social interactions, distinguished by the context of the involved parties, communication medium, or type of engagement e.g., direct vs. indirect, may influence the functioning of models, and for some tasks, such distinctions should be considered at the model design level. In our work, both the global interaction stream and local streams from the agents' perspective concern direct interactions.

The first perspective is dedicated to a single individual. It is a proposition for approximation of a single person as a data stream processing unit, when modelling larger society as a complex adaptive system. There are some strong similarities present when comparing a single person and a streaming algorithm from the angle of data processing restrictions and possibilities, that we described in the previous section. Both, people and streaming methods, have limited memory, unable to store a whole sequence of events they register. Both are required to react to the collected signals instantly. At last, both should be adaptable to change of the characteristics of inputs.
Collective systems are often studied with agent-based models where behaviour of a single agent (corresponding to an individual) is driven by a set of predefined, simple rules, and the social phenomena are observed as emerging from interactions between agents. Here, we propose that modelling those agents under the same conditions as streaming algorithms can bring them closer to reality. It could also automatically introduce some of the limitations related for instance with human cognitive processes based on streamingly arriving signals.

The second perspective is devoted to considerations on modelling a whole complex adaptive system, or at least its critical properties, with the use of data streams only. The classic approach for modelling collective adaptation processes requires building some kind of a social network as a social environment. As mentioned in Section 2, it is a tedious process that requires many informed decisions on the specific model of the network topology. Choice of an inadequate model might lead to wrong conclusions, as the social network structure has been demonstrated to vitally influence the course and outcomes of simulated processes~\cite{michalski2022temporal}. Specifically, there are situations in which building a social network is not possible due to computational limits, or the only source of information about the social system is an interaction stream. In that kinds of settings, a collective perspective is essential for modelling and analysis of complex social systems.

The absence of a social network structure representing social interactions poses significant challenge for modelling, as most of the well established models of social processes work on the basis of networks. Owing to that fact, research on new models or enhancements to the established ones is needed to provide the platform for modelling on top of interaction streams. Furthermore, it is not clear and yet to be studied what would be the capabilities of such models. Whether they are able to carry enough information to adequately represent social phenomena of interest.

\section{Social Learning Use Case}
\label{sec:social-learning}
Social learning is a complex phenomenon that can be considered an instance of collective adaptation framework. This approach focuses on how individuals within a society collectively adapt their knowledge based on the interactions with each other. Social learning models help to understand how groups form consensus, aggregate information to form beliefs and what are the conditions required for those to succeed. Data received by individuals in this concept come from observation and communication. Here, we would like to delve into details of how the streaming approach could be integrated into the modelling of the latter.

For social learning, similarly to other collective adaptation processes, the social environment, providing information on individuals sources of data, was represented with social networks. Both Bayesian~\cite{acemoglu2011bayesian} and non-Bayesian~\cite{degroot1974reaching,golub2010naive} models were designed to work in that regard. However, despite this fact, some of the social learning models include properties suited for streaming perspective at the single individual level. Bayesian agents, that rely their reasoning on Bayes' rule, process interaction events one by one and immediately update their internal state without keeping record of what happened in the past. Simpler models of opinion dynamics\cite{sznajd2002simple,clifford1973model,galam1986majority} also rely on the most recent interaction, thus the state of the individual can be determined by averaging it, based on their frequency or specific interaction properties.

On the other hand, there is no alternative approach for modelling social systems at the collective level with no network approach. The substantial difference between network-based and streaming approaches concerns the way that social interaction data are exploited. Network-based models firstly learn the social structure from data, usually as a fixed snapshot, or leverage some artificially generated topology that approximates a real-world setting. Then, an offline simulation is performed using the determined social network. The streaming approach, on the contrary, would rely on the real-world or generated social interaction data on the run.

Interaction stream based social learning models could be developed leveraging data stream processing techniques mentioned in Section 3, including both data-based techniques and task-based techniques. Another important aspect to consider is the observable set of stream record attributes. The simplest form of a data record, that we previously defined, assumes the presence of information about involved individuals' identity and ordering. The presence of additional attributes may allow for the development of more sophisticated models, capable of better representing social learning features, or doing it with greater precision. Furthermore, restrictions on memory size, available computational power and processing time are additional technical properties affecting models design and abilities.

The effects of selected methods, scope of interactions, and posed restrictions should be carefully examined. Preferably, network-based and data stream-based models should be compared using selected metrics and on artificially generated or real-world datasets with ground truth information about the social learning process. This kind of comparison would hopefully deliver information about limitations and precision gaps resulting from the streaming approach, as well as minimal conditions for it to succeed.

The fundamental question, whether it is possible to track social learning process effects and properties when working with interaction streams is yet to be answered. Unquestionably, monitoring the exact social traces of every individual in the observed community will not be possible, the way as it happens in network-based models, where the position and state of every agent in the simulation is known. Nevertheless, this feature is not a crucial component of an effective social learning model. In fact, the simulation results from agent-based models with stochastic components are usually averaged over number of runs, so a single simulation results do not provide any real value. The crucial information, that we seek to obtain, are the global features of the process at the level of a complete community e.g. the level of consensus over time, interactions category, or persistence of false beliefs~\cite{acemoglu2011opinion}.

\section{Conclusions and Outlook}
\label{sec:summary}
In this work we proposed a new perspective on modelling collective adaptation processes based on the social interaction streams that extends beyond the well-established network-based approach. We emphasized the similarities present between human cognitive processes and features of stream data mining algorithms, and proposed two modelling perspectives allowing to leverage the streaming approach for various collective processes. Finally, we presented a more detailed description of possible applications of streaming perspective to social learning use case.

We believe that future work in this matter should focus on exploring limitations of streaming methods on the ability to accurately representing social phenomena, compared the the network-based approaches. At this stage, we do not know how effectively interaction stream models can capture and predict group dynamics and decision-making processes in the context of social learning and overall collective behaviour. The aim of this work is to provide a background for future research in that direction. An in-depth analysis of stream preprocessing techniques is needed to answer the question of whether there are optimal and universal ones, applicable to the full range of specific social processes or rather they need to be tailored and carefully selected.

The last consideration is dedicated to the need for a more profound analysis of the interdisciplinary concept of social interactions. The advancements from fields of social psychology and neurobiology could provide additional insight into the computational perspective on interaction description and features.

Although this paper does not contain an experimental part, our aim in this perspective work was to provide an outlook on the essence of the stream-based modelling idea and prepare a proper justification for further research conducted in this direction. We hope it could serve as an encouragement to study social processes on top of social interaction streams, as this approach brings the models closer to the real-world settings and may be the only viable option for certain applications.

\subsubsection{Acknowledgements} This work was supported by the Polish National Science Centre, under grant no. 2021/41/B/HS6/02798. This work was also partially funded by the European Union under the Horizon Europe grant OMINO (grant no. 101086321). Views and opinions expressed are however those of the author(s) only and do not necessarily reflect those of the European Union or the European Research Executive Agency. Neither the European Union nor European Research Executive Agency can be held responsible for them.

\bibliographystyle{splncs04}
\bibliography{references}
\end{document}